\documentclass{article}
\usepackage{latexsym}
\usepackage[psamsfonts]{amssymb}
\usepackage{amsmath}

\def\QED{\hskip0.1em\hfill\null\ \null\nobreak\hfill\kern3pt\vbox{\hrule\hbox
   {\vrule\kern1pt\vbox{\kern1.7pt\hbox{$\scriptscriptstyle{QED}$}
    \kern0.2pt}\kern1pt\vrule}\hrule}}

\def\END{\hskip0.1em\hfill\null\ \null\nobreak\hfill\kern3pt\vbox{\hrule\hbox
   {\vrule\kern1pt\vbox{\kern1.7pt\hbox{$\,\,\,\vspace{5pt}$}
    \kern0.2pt}\kern1pt\vrule}\hrule}}
\newtheorem{theorem}{Theorem}
\newtheorem{lemma}{Lemma}
\newtheorem{corollary}{Corollary}
\newtheorem{proposition}{Proposition}
\newtheorem{remark}{Remark}
\newtheorem{definition}{Definition}
\newtheorem{example}{Example}

\newcommand{\bCd}{\bEq\begin{CD}}
\newcommand{\eCd}{\end{CD}\eEq}
\newcommand{\bcd}{\beq\begin{CD}}
\newcommand{\ecd}{\end{CD}\eeq}
\newcommand{\ben}{\begin{enumerate}}
\newcommand{\een}{\end{enumerate}}
\newcommand{\bEq}{\begin{eqnarray}}
\newcommand{\eEq}{\end{eqnarray}}
\newcommand{\beq}{\begin{eqnarray*}}
\newcommand{\eeq}{\end{eqnarray*}}
\newcommand{\bDf}{\begin{definition}\em}
\newcommand{\eDf}{\end{definition}}
\newcommand{\bLm}{\begin{lemma}}
\newcommand{\eLm}{\end{lemma}}
\newcommand{\bPr}{\begin{proposition}}
\newcommand{\ePr}{\end{proposition}}
\newcommand{\bTh}{\begin{theorem}}
\newcommand{\eTh}{\end{theorem}}
\newcommand{\bCr}{\begin{corollary}}
\newcommand{\eCr}{\end{corollary}}
\newcommand{\bRm}{\begin{remark}\em}
\newcommand{\eRm}{\end{remark}}
\newcommand{\bEx}{\begin{example}\em}
\newcommand{\eEx}{\end{example}}

\newcommand{\ie}{{\em i.e$.$} }
\newcommand{\eg}{{\em e.g$.$} }

\newcommand{\A}{\forall}

\newcommand{\cD}{\mathcal{D}}

\newcommand{\bp}{\boldsymbol{p}}

\newcommand{\bz}{\boldsymbol{z}}

\newcommand{\bE}{\boldsymbol{E}}

\newcommand{\bG}{\boldsymbol{G}}

\newcommand{\bP}{\boldsymbol{P}}

\newcommand{\bV}{\boldsymbol{V}}

\newcommand{\bZ}{\boldsymbol{Z}}

\newcommand{\tht}{\theta}

\newcommand{\lam}{\lambda}

\newcommand{\ome}{\omega}

\DeclareFontFamily{U}{UWCyr}{}
\DeclareFontShape{U}{UWCyr}{m}{n}{%
    <5> <6> <7> <8> <9>
    <10> <10.95> <12> <14.4> <17.28> <20.74> <24.88> wncyr10
    }{}
\DeclareMathAlphabet{\cyrm}{U}{UWCyr}{m}{n}

\def\con{{\offinterlineskip\lower 1truept\hbox{\kern2pt
\vbox to7truept{\vfill\hbox to4truept{\hrulefill}}\vrule \kern3pt}}}

\newcommand{\bdg}{\begin{diagram}}
\newcommand{\edg}{\end{diagram}}

\title{{\Large {\bf 
Algebraic structures generating reaction-\\diffusion models:  the activator-substrate system}}}

\author{Marcella Palese \\
{\footnotesize Department of Mathematics, University of Torino} \\ {\footnotesize via C. Alberto 10, I-10123 Torino, Italy } \\ 
 {\footnotesize {\sc e-mail: marcella.palese@unito.it}}}

\date{}

\begin{document}
\maketitle

\begin{abstract}{\small

We shall construct
a class of nonlinear reaction-diffusion equations 
starting from an infinitesimal algebraic skeleton.
Our aim is to explore the possibility of an algebraic foundation of integrability properties and of stability of equilibrium states associated with nonlinear models describing patterns formation.}


\end{abstract}

 \section{Introduction}

In his famous paper, Turing suggested that a system of chemical substances reacting together and diffusing through a tissue, could describe the main phenomena of morphogenesis \cite{Tur52}. In particular, in his work it was emphasized that patterns could appear if one of the substances diffuses much faster than the other.

Nonlinear reaction-diffusion systems have then been proposed to answer the question about how cells, under the influence of their common genes, could produce spatial patterns, see \eg \cite{KoMe94} and references therein. 
They consist in models describing generation of patterns from an initially homogeneous state taking into account the relevance of chemical gradient in biological systems, in particular nonlinear interaction of two chemicals and their diffusion.
Patterns formation turns out then to be the output of {\em local self-enhancement} such as local autocatalysis and {\em long range  inibition}.
A simple model proposed by Koch \& Meinhardt is the activator-substrate systems (constituted by a self-enhanced reactant and a depleted reactant which plays the role of the antagonist).
In the simplest mathematical form, only few relevant parameters characterize the model: the normalized {\em diffusion constant} and the normalized {\em  cross-reaction coefficient}.
In particular, the inibition due to the substrate reactant can be effective if the normalized diffusion constant is $<< 1$ (that means the diffusion constant of the activator should be much lesser than the diffusion constant of the substrate), a necessary condition for the generation of stable patterns.

Recent studies  pointed out how, beside being a universal principle explaining regular pattern formation in chemical, physical and biological morphogenesis systems,  Turing's activator-inibitor principle is also at the basis of regular pattern formation in a variety of ecosystems. 
In particular, the scale-depending feedback is a unifying ecological principle playing a key r\^ ole and mainly consisting  in short-range ecological facilitation such as local modification of the environment and long-range competition for resources; see \eg \cite{RiKo08} whereby {\em the prerequisite of long-distance negative feedback is proposed as a basic principle for regular pattern formation in ecosystems}.

In this paper we show that this condition can be recognized as a condition on the internal symmetry properties of a system. This perspective, concerned with the action of a group of transformations on the space of  possible configurations (`fields'), will be exploited in Section \ref{Section 2} and detailed in the Appendices. 

In particular, in Section \ref{Section 3} we shed new light on the long-range negative feedback condition by a case study exploiting the internal symmetry algebra
of {\em twisted} reaction-diffusion equations {\em with a null basic production term}. 
We observe that, in the limit of a null normalized diffusion constant, such a condition is related with the existence of soliton solutions - \ie  travelling waves remaining stable after interactions.
We can see that requiring the normalized diffusion constant  being null corresponds to look for specific symmetry properties of the associated system; indeed this integrable case can be seen as a limiting case of the internal symmetry algebra (see  Appendix 2).

As a major result, in subsection \ref{main} we recover {\em activator-substrate systems} by performing a slight modification of the internal symmetry algebra of twisted reaction-diffusion equations. We obtain as a byproduct that the necessary condition for the generation of stable patterns for such a system, besides being  related with general integrability properties in the limit of a null normalized diffusion constant, can also be  formulated in terms of `closeness' properties within the symmetry algebra vector space.

This moreover suggests the possibility to enhance the interplay of the two different approaches to the study of stability: the qualitative study  based on the analysis of the stability of equilibrium configurations and global integrability properties related with existence of stable solutions such as travelling waves or solitons.

The scope of this work is therefore to propose an intrinsic algebraic form of the local mechanisms expressed by reaction-diffusion partial differential equations. 
Our considerations are based on the well known duality between linear differential forms and tangent vector fields on manifolds, and make use of algebraic and geometric techniques developed within a theoretical physics framework, see \eg 
\cite{Es82,PaWi02,PaWi10,PaWi11,PRS79}, in particular suitable generalization of the structure equations of a Lie group.
In the present note, we show that the appearing and the significance of parameters in a model can be characterized within an algebraic-geometric formulation in terms of integrable ``towers with infinitesimal algebraic skeletons'' (see Appendix 1), which generalizes that of integrable connections on spaces of configurations with symmetries.

\section{Underlying algebraic skeletons}\label {Section 2} 

Concerning real ecosystems in homogeneous landscapes, establishment and survival of organisms can be inhibited by limited resources or other stress factors: it is an empirical evidence that, although the organisms can acts only locally, the effects have influence at distance and therefore are of global nature, showing a dependence  on density of the organisms themselves \cite{RiKo08}.

In this paper we propose an approach which is based on the study of global properties of partial differential equations such as internal symmetries and invariance properties having however an issue in dynamics. The underlying idea is that transformations of configurations of a system can be globally studied  by means of the theory of the action of Lie groups on manifolds. Differential equations therefore are an issue of the differential content carried by a Lie group (and its Lie algebra) and by its structure equations providing connections on the space of configurations. 

Although such mathematical tools could appear of a quite abstract nature, nevertheless they are the right tools to deal with global properties at large scales, being natural tools for connecting local data to global ones. 
In particular they implement the concept of `positional information' \cite{Wol69}. In Appendix 1 the general context is sketched and appropriate references to the abstract theory are given.

It is well known that, looking for invariance properties, a ``prolongation'' algebra can be associated with a given system of nonlinear partial differential equations. Nonlinearity, then,  results in a quite intriguing algebraic structure which is  ``only partially'' a Lie algebra (we speak of an `open' Lie algebra structure)  \cite{WaEs75}.
By an inverse procedure based on the intrinsic duality between Lie algebras and differential systems \cite{Es82}, open Lie algebraic structures can `generate' whole families of different nonlinear systems bound by the same internal symmetry structure. 

We show that a slight modification of the internal symmetry properties generates {\em new} models which can contain possible integrable subcases. 
As an example, by a modification of the symmetry algebra associated with a model for pattern formation on the shells of molluscs \cite{MeKl87} we shall recover the activator-substrate reaction-diffusion model proposed by Koch \& Meinhardt.
In both models the necessary condition for the generation of stable patterns, \ie the prerequisite of long-distance negative feedback, can be interpreted already at a symmetry level, as a ``closeness condition'' in  the symmetry space (see the concluding Remark).

Our results are based on the observation that two fundamental aspects are involved in pattern formations: symmetries on the one side (algebraic content) and changes in time and space on the other side (differential content).
In particular, to keep account of the `interaction' of both aspects, we act effectively with open Lie algebraic structures by using their refined structure: we need to introduce, 
a notion which generalizes the concept of  a homogeneous space, \ie that of  an {\em algebraic skeleton} $\bE = \mathfrak{g} \oplus \bV$ on a finite-dimensional vector space $\bV$, with $\mathfrak{g} $ a possibly infinite dimensional Lie algebra (see Appendix  1).

Consider then the following infinite dimensional vector space $\bE =\mathfrak{g} \oplus  \bV$
\beq
& & [\psi_1,\psi_2] = 0, [\psi_1,\psi_3] = 0, [\psi_1,\psi_4] = 0,  [\psi_1,\psi_5] = ?, 
\\
& & [\psi_1,\psi_6] = 0, [\psi_1,\psi_7] = 0, [\psi_1,\psi_8] = 0 \,   
\cdots 
\\
& & 
  [\psi_2,\psi_3] = -2\cD  \psi_6, [\psi_2,\psi_4] = 2\cD \psi_7,  [\psi_2,\psi_5] = ?, [\psi_2,\psi_6] = - 2 \cD \psi_2, 
  \\
& & [\psi_2,\psi_7] = 0, [\psi_2,\psi_8] = [\psi_4,\psi_6] + [\psi_3,\psi_7] \, 
  \cdots \\
& & 
  [\psi_3,\psi_4] = 2\cD \psi_8,  [\psi_3,\psi_5] = ?, [\psi_3,\psi_6] =  2 \kappa \cD \psi_3, 
  \\
& & [\psi_3,\psi_7] = [\psi_2,\psi_8] - [\psi_4,\psi_6], [\psi_3,\psi_8] = 0 \, 
\cdots 
\\
& &    [\psi_4,\psi_5] = 0, [\psi_4,\psi_6] =   \cD[\psi_1,\psi_5], [\psi_4,\psi_7] = [\psi_2,\psi_5]  - 2\cD\psi_2, 
\\
& & [\psi_4,\psi_8] = \cD [\psi_3,\psi_5]  \, 
  \cdots \\
& & 
 [\psi_5,\psi_6] =  [\psi_2, [\psi_3,\psi_5] ]  - [\psi_3, [\psi_2,\psi_5] ]  , [\psi_5,\psi_7] =  [\psi_4, [\psi_2,\psi_5] ] , 
 \\
& &  [\psi_5,\psi_8] =  [\psi_4, [\psi_3,\psi_5] ]   \, 
 \cdots 
 \\
& &  
  [\psi_6,\psi_7] = 2\cD\psi_7 - [\psi_1,[\psi_2,\psi_5] ] ,  [\psi_6,\psi_8] = -2\kappa \cD\psi_8 - \cD  [\psi_3,[\psi_1,\psi_5] ]   \, 
  \cdots
  \\
& & 
 [\psi_7,\psi_8]  = [\psi_4, [\psi_3,\psi_7] ] - [\psi_3, [\psi_4,\psi_7] ] \,  \cdots \,
\eeq  
where $\cD$,  $\kappa$ are real parameters and   `?' denotes undefined commutators.

Note that the commutators $[\psi_1,\psi_5] $, $[\psi_2,\psi_5]$ , $[\psi_3,\psi_5] $ are not defined and in particular that even  introducing new generators $\psi_9,\psi_{10},\psi_{11}$, the algebra anyway does not close as a Lie algebra and is therefore an `open' Lie algebra structure. Let us describe its refined structure. To this aim, 
it is also important to stress that many other brackets between elements of the above vector space $\bE$ are not defined as Lie algebra brackets since they are given in terms of $[\psi_1,\psi_5] $, $[\psi_2,\psi_5]$, $[\psi_3,\psi_5] $. 

The vector space $\bV$ is finite dimensional and generated by $[\psi_1,\psi_5]$, $[\psi_2,\psi_5]$, $[\psi_3,\psi_5]$, $\psi_5$. It has the property that each  bracket of $\psi_1, \psi_2, \psi_3, \psi_4, \psi_6,\psi_7,\psi_8$ (freely generating an infinite dimensional Lie algebra $\mathfrak{g}$) with its generators is again in $\bV$. The Lie algebra $\mathfrak{g}$ can be characterized as a Lie algebra of invariant vector fields tangent to the fibers of a bundle with structure group $G$; therefore $\psi_i$ can be regarded as linear differential operators by the correspondence
$
\psi_i \rightarrow \psi_i^{j}\frac{\partial}{\partial \xi^{j}}$, 
while the Lie brackets can be interpreted as commutators $[\psi_i^{j}\frac{\partial}{\partial \xi^{j}}, \psi_k^{l}\frac{\partial}{\partial \xi^{l}} ]=\psi_i^{j}\frac{\partial}{\partial \xi^{j}} \psi_k^{l}\frac{\partial}{\partial \xi^{l}} -  \psi_k^{l}\frac{\partial}{\partial \xi^{l}}\psi_i^{j}\frac{\partial}{\partial \xi^{j}}=
c_{ik}^m  \psi_m^{n}\frac{\partial}{\partial \xi^{n}}$, with $c_{ik}^m$ the structure constants of $\mathfrak{g}$.

In particular, we note that the commutator relations 
$[\psi_3,\psi_7]  - [\psi_2,\psi_8] = \cD[\psi_1,\psi_5] $, 
$ [\psi_6,\psi_7] - 2\cD\psi_7= [\psi_1,[\psi_2,\psi_5] ] $,  
$[\psi_6,\psi_8] + 2\kappa \cD\psi_8 = \cD  [\psi_3,[\psi_1,\psi_5] ] $, 
$[\psi_4,\psi_7]  + 2\cD\psi_2 = [\psi_2,\psi_5]$ and the related 
$[\psi_7,\psi_8]  = [\psi_4, [\psi_3,\psi_7] ] - [\psi_3, [\psi_4,\psi_7] ]$ 
say that unknown commutators in the freely generated  Lie algebra are related in such a way that their assigned relations are elements of $\bV$; according to Appendix 1,  {\em the above algebraic relations define an infinitesimal skeleton}. 

It is important to understand that such a feature of the  commutator relations {\em constitutes a skeleton's peculiar property  which generates nonlinearity}. Nonlinear terms appearing  in the `generated' differential systems are dual to such relations, therefore {\em commutator relations of   skeletons are the algebraic counterpart of nonlinear population interactions}.

\subsection{Models for pattern formation on the shells of molluscs}

In order to exploit symmetry properties, we can define an homomorphism of the infinite dimensional  freely generated  Lie algebra with a quotient Lie algebra by fixing unknown commutators; we can do this either by  fixing the value of $[\psi_1,\psi_5]$, $[\psi_2,\psi_5]$, $[\psi_3,\psi_5]$ as generated by elements in the freely generated  Lie algebra, or by fixing the value of the unknown commutators 
$[\psi_2,\psi_8]$,
$[\psi_3,\psi_7]$,
$[\psi_4,\psi_7]$,
$[\psi_6,\psi_7]$,
$[\psi_6,\psi_8]$.

Details on various choices can be found in  Appendix 2. Here we point out the case
obtained by 
putting  $\psi_7= 0$  and  $\psi_6 = 0$ (since $\psi_3\neq 0$ and $\cD\neq 0$ this implies $\kappa=0$, \ie the case where one of the reaction coefficient vanishes)
we obtain an homomorphism with the Lie algebra corresponding to a group of Euclidean movements in the plane 
\beq
[\psi_1,\psi_2] = 0,   [\psi_1,\psi_5] = 0, [\psi_2,\psi_5] =  2\cD \psi_2\,.
\eeq

The above results seem to be in agreement with what announced concerning the case $\cD\geq 0$ in \cite{MPS97}, whereby, instead, integrability of reaction-diffusion type equations with $\cD < 0$, \ie diffusion constants of opposite sign, have been studied (such systems play 
 a r\^ole in gauge theory of gravity \cite{MPS97}; similarities with coupled nonlinear Schroedinger equations describing the waves propagation in optical fibres can be recognized, see \eg \cite{PaWi11}; both cases are integrable and homomorphisms with  infinite dimensional loop Lie algebras have been determined).  

However, let us stress that alternatively  we can choose 
\beq
[\psi_1,\psi_5] = \lam \psi_1,  [\psi_2,\psi_5] = \lam \psi_3,  [\psi_3,\psi_5] =0
\eeq
which define an homomorphism with a finite dimensional Lie algebra with spectral parameter $\lam$. 

This latter case implies $\cD=0$, \ie the diffusion constant of the activator is zero;  {\em the system would be integrable and would admit a Lax pair}. In fact, the case $\cD=0$ has been related with existence of travelling waves; it appears in a model for pattern formation on the shells of molluscs \cite{MeKl87} and {\em it is perhaps important to note that it is a limit case of $\cD<<1$}. In  the following we shall consider the case $0 < \cD<<1$ and we shall obtain the Koch \& Meinhardt  activator-substrate system from an integrability condition for a tower with a slightly modified skeleton.

\section{Reaction-diffusion models from integrable towers with skeleton $\bE$}\label{Section 3}

We shall now explain {\em how skeletons can `generate' nonlinear differential systems}.

\subsection{Twisted reaction-diffusion models}

For the purpose of this work, in fact, it is now a very remarkable feature that it is possible to obtain reaction-diffusion type models directly from the skeleton by using a generalization (a truncated version) of the structure equations. Indeed we need to introduce a way to produce (exterior) differential equations. The differential structure can be modelled on the assigned skeleton by its absolute parallelism forms (see Appendix 1); the corresponding integrability conditions are given by 
\beq
\omega^k=d\xi^k-\psi_j^k\theta^j \,,  \qquad  d\omega^k=\psi_j^kd\theta^j-{1\over 2}[\psi_j,\,\psi_i]^k\theta^j\wedge
\theta^i = 0 \,, \quad ({\rm mod}\;\omega^k) \,;
\eeq
where $\tht^k$ are some linear differentials (horizontal $1$-forms on $\bP\to\bZ$) and $\wedge$ stands for the skew (also called exterior) product.

This provides us with the differential  content we are looking for.
In fact, the forms $\ome^k$ can be recognized as tower forms of Cartan type (\ie as a pull-back of contact forms by a B\"acklund map \cite{PaWi02}) if the following exterior differential contraints are satisfied  by the $\tht^k$:
\beq
& & d\theta^1=0,  
\quad d\theta^4 = 0,
\quad d\theta^5 = 0,
\\
& &
d\theta^2 - 2 \cD \theta^2\wedge\theta^6 - 2 \cD \theta^4\wedge\theta^7= 0,  
\quad d\theta^3  + 2\kappa \cD\theta^3\wedge\theta^6 = 0,
\\
& &
d\theta^6-\cD \theta^2\wedge\theta^3 = 0,
\quad d\theta^7+\cD \theta^2\wedge\theta^4+ 2 \theta^6\wedge\theta^7= 0,
\\
& &
d\theta^8+\cD \theta^3\wedge\theta^4 + 2\kappa\cD \theta^6\wedge\theta^8 = 0, \\
& &
\theta^1\wedge\theta^5 + \cD\theta^2\wedge\theta^8 + \cD \theta^4\wedge\theta^6 = 0, 
\quad \theta^2\wedge\theta^5+\theta^4\wedge\theta^7= 0,
\\
& &
 \theta^2\wedge\theta^8+\theta^3\wedge\theta^7 = 0,
\quad \cD\theta^3\wedge\theta^5 - \theta^4\wedge\theta^8 = 0,
\\
& &
\theta^5\wedge\theta^6 = 0, 
\quad \theta^5\wedge\theta^7 = 0,
\quad \theta^5\wedge\theta^8 = 0,
\quad  \theta^6\wedge\theta^8 = 0, 
\quad \theta^7\wedge\theta^8 = 0 \,.
\eeq 
A solution is given by:
\beq
&& \theta^1 = dt, 
\quad \theta^5 = \frac{1}{\cD} dt, 
 \quad \theta^6 = -\frac{1}{\cD}\mu\nu dt, 
  \quad \theta^7 = \frac{1}{\cD}\mu dt, 
  \\
  && \theta^8 =- \frac{1}{\cD}\nu dt, 
    \quad  \theta^4 = \frac{1}{\cD} dx, 
     \quad \theta^2 = \frac{1}{\cD}(\mu dx + \mu_x dt), 
      \quad \theta^3 = -\frac{1}{\cD}(\nu dx + \nu_x dt) \,,
\eeq
where $\mu$ and $\nu$ are
 functions,
 depending on $x$ and $t$, 
which must satisfy the following {\em twisted} reaction-diffusion equations
\beq
\mu_{ t} - \cD \mu_{xx} - 2 {\mu}^2\nu- 2\mu = 0, 
\qquad 
\nu_{t} - \nu_{xx} + 2 \kappa \mu{\nu}^2 = 0\,;
\eeq
here, as usual, the  supscripts means partial derivatives.

This system can be recognized as {\em a twisted  reaction-diffusion model 
with a zero basic production term}, with $\cD$ a normalized diffusion constant and $ \kappa$ a normalized cross-reaction coefficient.
It {\em contains}, as a limit feature, the particular case $\cD=0$ of a model for pattern formation on the shells of molluscs \cite{MeKl87}.

\subsection{The activator-substrate reaction-diffusion model} \label{main}
In \cite{RiKo08} it  has been pointed out that various examples of real ecosystems, such as arid, wetland, savanna ecosystems, coral reefs, mussel beds, {\em etc.}  can be described as activator-(depleted)substrate systems \cite{GiMe72,KoMe94,Me95}.  
In the following we shall obtain such type of systems by acting on the skeleton of the twisted reaction-diffusion system above.

As we already explained the differential structure (\ie the absolute parallelism) is totally of general nature, while the somewhat `true' content of a specific model comes from the algebraic stuctures we insert in the structure equations. This fact suggest the possibility of characterizing different models by their algebraic content. 

Let us now, in fact, consider a slight change in the algebraic skeleton, such as 
\beq 
[\psi_2,\psi_6] = 2\kappa \cD\psi_3 - 2 \cD \psi_2,  \quad  [\psi_3,\psi_6] = 0, \quad   [\psi_4,\psi_5] =  2\kappa \cD \psi_3\,.
\eeq
Such a change provides the exterior differential equation 
\beq
d\theta^3  - 2\kappa \cD\theta^2\wedge\theta^6 - 2\kappa \cD\theta^4\wedge\theta^5= 0
\,,
\eeq
and therefore originates the system:
\beq
\mu_{ t} - \cD \mu_{xx} - 2(\nu \mu^2- \mu) = 0, 
\qquad 
\nu_{t} - \nu_{xx} - 2 \kappa(1- \nu\mu^2) = 0\,. 
\eeq
\ie the activator-substrate reaction-diffusion model proposed by Koch \& Meinhardt (in a more general form previously also appeared in Gierer \& Meinhardt \cite{GiMe72}); here again $\cD$ is a normalized diffusion constant and $ \kappa$ a normalized cross-reaction coefficient.

\bRm \label{Remark}
We stress that the latter system has been obtained by operating a slight change in the algebraic skeleton of the former and it is rather  different from that one. 
Note, in particular, that activator-substrate systems have the property that both time derivatives $\mu_{ t}$ and $\nu_{ t}$ (activator production term and depletion term) are proportional to $\mu^2$ and depend linearly on $\nu$, while the system obtained in the subsection above is in a {\em symmetrically  twisted form}. 

We see, indeed, that commutator relations of the skeleton are the algebraic counterpart of nonlinear population interactions. This implies that {\em the parameters appearing in a given model, and even the form of the model itself}, can be somewhat `controlled' already at an algebraic level. This is in agreement with the fact that, while Lie group actions provide algebraic forms of dynamics, `deformations' of Lie algebraic structures, such as skeletons, provide the nonlinear content.

Moreover, suppose a metrics could be defined on $\bE$, so that one can reasonably think of a condition of closeness, which we write for simplicity as $
[\psi_4,\psi_7] $ $ \simeq $ $ [\psi_2,\psi_5]$.
Having a glance at the skeleton structure, it is evident that such a condition would be equivalent to the request that $\cD << 1$, 
\ie the diffusion constant of the activator be much lesser than the diffusion constant of the substrate; therefore, we can characterize a condition for the appeareance of patterns by means of properties of vectors  in $\bE$.

Finally, in order to highlight and emphasize the significance of our approach, let us stress that the latter statement is a global symmetry-related expression of (and are in agreement with) empirical observations. In fact,  beside the necessity that organisms modify their environment by inducing a long-range negative feedback thus allowing regular ecological pattern formations, the strenght of this feedback depends on the density of the organisms at large scale \cite{RiKo08}.
Indeed, approaches based only on the actual differential expression of a system, necessarily of a local nature, could overlook aspects related to {\em global 
 properties} of the system itself such as for example, internal symmetries.
\eRm

\section{Conclusions}

Our results are of general nature and in principle could be applied to other mathematical models proposed in various branches of biology and ecology, see \eg \cite{Murray} and, for a review and further developments, \cite{VoPe09}; of particular interest would be the possibility of  application to models with delay. In fact, spatio-temporal pattern formation can
be caused by time delay factors. The study of algebraic structures generating models with delay would be therefore of particular interest in comparing at an intrinsic algebraic level the various approaches in modeling pattern formation. 

Moreover, according to the Remark above, we enhance the possibility of an algebraic-geometric study of the stability of the equilibrium states.
In particular the emerging of both Hopf and Turing bifurcations depends on the parameters range and their mutual relations in a given model: we saw that such characters could be formalized already at the algebraic level in terms of the representation of $\mathfrak{g}$ on $\bV$ (commutator relations of skeletons).

\section*{Acknowledgements}
Research supported by Department of Mathematics University of Torino through local research project {\em Metodi Geometrici in Fisica Matematica e Applicazioni 2013-2015}.

\section*{Appendix 1}\label{1}

\subsection*{Towers with skeletons}

Let us first recall a few mathematical tools constituting the background for a detailed treatment of which 
we refer to \cite{PaWi02,PaWi10,PaWi11} and \cite{Mo93,PRS79}; the use of the concept of tower with skeleton has been inspired by (and is a mathematical generalization of) the procedure outlined in \cite{LeSo89}.

An {\em algebraic skeleton} on a finite-dimensional vector space $\bV$ is a triple $(\bE,\bG,\rho)$, with $\bG$ a (possibly infinite-dimensional) Lie group, 
$\bE = \mathfrak{g} \oplus \bV$ is a (possibly infinite-dimensional) vector space {\em not necessarily equipped with a Lie algebra structure}, 
$\mathfrak{g}$ is the Lie algebra of $\bG$, 
and 
$\rho$ is a representation of $\mathfrak{g}$ on $\bE$ such that it reduces to the adjoint representation of $\mathfrak{g}$ on itself. The fact that $\bE$ is not a direct sum of Lie algebras, but an open algebraic structure is fundamental in order to be able to generate whole families of nonlinear differential systems, starting from it.

We now consider a suitably constructed differentiable structure which is somewhat modelled on the skeleton above. Let us introduce a differentiable manifold $\bP$ on which a Lie group $\bG$, with Lie algebra $\mathfrak{g}$, acts on the right; $\bP$ is a principal bundle $\bP\to\bZ\simeq\bP/\bG$. 
By construction, we have that $\bZ$ is a manifold of type
$\bV$, \ie $\A \bz\in \bZ$,  $T_{\bz} \bZ \simeq \bV$. Suppose we have a way to define a representation $\rho$ of the Lie algebra $\mathfrak{g}$ on $T_{\bz} \bZ \simeq \bV$, in such a way that it could be possible under certain conditions to find a %
 homomorphism between the open infinite dimensional Lie algebra, constructed by $\rho$, and a quotient Lie algebra.
Let us call $\mathfrak{k}$ the (possibly infinite dimensional) Lie algebra obtained as the direct sum of  such a quotient Lie algebra  with $\mathfrak{g}$. 
From the differentiable side, a {\em tower $\bP(\bZ,\bG)$ on $\bZ$ with skeleton $(\bE, \bG,\rho)$ } is 
an {\em  absolute parallelism} $\ome$ on $\bP$ valued in $\bE$, 
invariant with respect to $\rho$ and reproducing elements of $\mathfrak{g}$ from the fundamental
vector fields induced on $\bP$. 
Let then $\mathfrak{k}$ be a Lie algebra and $\mathfrak{g}$ a Lie subalgebra of
$\mathfrak{k}$. 
Let $\bG$ be a Lie group with Lie algebra $\mathfrak{g}$ and $\bP(\bZ, \bG)$ be a principal fiber bundle with structure group $\bG$ over a manifold $\bZ$ as above. 
A {\em Cartan connection} in $\bP$ of type
$(\mathfrak{k}, \bG)$ is 
a $1$--form $\ome$ on $\bP$ with values in
$\mathfrak{k}$ such that
$\ome |_{T_{\bp} \bP}: T_{\bp} \bP\to \mathfrak{k}$ is an isomorphism $\forall \bp \in
\bP$, 
$R^{*}_{g}\ome=Ad(g)^{-1}\ome$ for $g\in \bG$ 
and 
reproducing elements of $\mathfrak{g}$ from the fundamental
vector fields induced on $\bP$.
It is clear that a Cartan connection
$(\bP, \bZ, \bG, \ome)$ of type $(\mathfrak{k}, \bG)$ is a special case of a tower on $\bZ$.
In the following, we shall be interested in the case when from a tower one can construct a Cartan connection by a quotienting.

\section*{Appendix 2}\label{2}

\subsection*{Homomorphisms with finite dimensional Lie algebras} 

We can find a  homomorphism with a finite dimensional Lie algebra from the infinite dimensional open Lie algebra generated by  the skeleton $\bE$ given in Section \ref{Section 2} 
by taking the following quotient, with $\lam$ being a real parameter.
\beq
 [\psi_3,\psi_7] = -\lam \psi_4\,, \quad [\psi_3,\psi_4] = \lam \psi_7 
 \,, \quad [\psi_4,\psi_7] = \lam \psi_3 
 \,.
\eeq
We get then $\lam \psi_7 = 2\cD\psi_8$,  
$[\psi_4,\psi_6] = \lam \psi_4$,
and, provided that $\cD \neq 0$,
\beq
[\psi_1,\psi_5] = \frac{\lam}{\cD}\psi_4  \,, \quad [\psi_2,\psi_5] = 2\cD\psi_2+\lam \psi_3\,, \quad [\psi_3,\psi_5] = \frac{\lam}{\cD}\psi_3\,.
\eeq
Since we also have that $[\psi_4,\psi_8] = \cD [\psi_3,\psi_5]$ (see the skeleton), for the consistency of the relations, we get in particular
$
\lam = 2\cD$; 
 thus we can write 
\beq
 [\psi_7,\psi_3] = 2\cD \psi_4\,, \quad [\psi_3,\psi_4] = 2\cD \psi_7 
 \,, \quad [\psi_4,\psi_7] = 2\cD \psi_3 
 \,.
\eeq
It is easy to see that in this case the algebra closes as a Lie algebra
\beq
[\psi_1,\psi_5] = 2\psi_4  \,, \quad [\psi_2,\psi_5] = 2\cD(\psi_2+ \psi_3)\,, \quad [\psi_3,\psi_5] = 2\psi_3\,.
\eeq
Furthermore, since $\psi_7 =  \psi_8$, then  $[\psi_6,\psi_8] = -2\kappa \cD\psi_8 - \cD  [\psi_3, [\psi_1,\psi_5] ] = [\psi_6,\psi_7] = 0$ provides, by an iterated application of the Jacobi identity, the condition $-2\kappa \cD +2 \cD=0$. Then, if $\cD\neq 0$,  we must have $\kappa = 1$, corresponding to activator and substrate having the same cross-reaction coefficients. Therefore we get
\beq
[\psi_2,\psi_3] = -2\cD  \psi_6 \,, \quad [\psi_3,\psi_6] =  2 \cD \psi_3\,, \quad [\psi_2,\psi_6] = - 2 \cD \psi_2\,;
\eeq
but  the consistency of $[\psi_5,\psi_7] =  [\psi_4, [\psi_2,\psi_5] ]$ and $[\psi_5,\psi_8] =  [\psi_4, [\psi_3,\psi_5] ]$ implies  $\cD = \frac{1}{2}$, which corresponds to a closed Lie algebra (without a spectral parameter).

Let us then consider a different closing homomorphism by setting $\psi_8 =0$ and  $[\psi_3,\psi_7] $ $=$ $ -\lam \psi_4$, $ [\psi_4,\psi_7] $ $=$ $  \lam \psi_3$.
We get  $[\psi_3,\psi_5]=0$,  
$[\psi_4,\psi_6] = \lam \psi_4$,
and, in particular, provided that $\cD \neq 0$,
\beq
[\psi_1,\psi_5] = \frac{\lam}{\cD}\psi_4  \,, \quad [\psi_2,\psi_5] = 2\cD\psi_2+\lam \psi_3\,, \quad [\psi_3,\psi_5] = 0\,.
\eeq
By the Jacobi identity, from $[\psi_5,\psi_7] = -4\cD^2\psi_7$, and $[\psi_1,\psi_7] = 0$, we get $ [\psi_4,\psi_7] =0$ which implies $\lam\psi_3 =0$. We esclude the trivial case $\psi_3 = 0$;  therefore we let  $\lam=0$. 
In this case the commutation relations become
$ [\psi_3,\psi_7] = 0$, $ [\psi_3,\psi_4] =0$,
$[\psi_4,\psi_7] = 0 $, $[\psi_1,\psi_5] =0 $, $[\psi_2,\psi_5] = 2\cD\psi_2$,  $[\psi_3,\psi_5] = 0$; the remaining commutators are
\beq
&[\psi_5,\psi_7] = -4\cD^2\psi_7 \,, \quad [\psi_5,\psi_6] = -4\cD^2\psi_6 \,, \quad [\psi_6,\psi_7] = 2\cD\psi_7\,.
\\
&[\psi_2,\psi_3] = -2\cD  \psi_6 \,, \quad [\psi_3,\psi_6] =  2 \kappa\cD \psi_3\,, \quad [\psi_2,\psi_6] = - 2 \cD \psi_2\,,
\eeq
while we do not get a constraint on $\kappa$.


\end{document}